\documentclass[usenatbib]{mn2e}

\usepackage[dvips]{epsfig,color}

\usepackage{amssymb,amsmath}

\newcommand{\eg}{e.g.}
\newcommand{\ie}{i.e.}
\newcommand{\etal}{et al.}

\newcommand{\ignore}[1]{\relax}

\DeclareMathAlphabet{\mathsfsl}{OT1}{cmss}{m}{sl}
\DeclareMathOperator{\sech}{sech}

\newcommand{\mt}{\ensuremath{M}}
\newcommand{\fst}{\ensuremath{\tilde f}}
\newcommand{\rhot}{\ensuremath{\tilde \lambda}}

\newcommand{\dif}{\mathrm{d}}

\begin{document}

\title{Dynamics of One-dimensional Self-gravitating Systems Using
Hermite-Legendre Polynomials}
\author[Barnes \& Ragan]{
Eric I. Barnes\thanks{email:barnes.eric@uwlax.edu}, Robert J.
Ragan\thanks{email:rragan@uwlax.edu} \\
Department of Physics, University of Wisconsin --- La
Crosse, La Crosse, WI 54601}

\maketitle

\begin{abstract}

The current paradigm for understanding galaxy formation in the
universe depends on the existence of self-gravitating collisionless
dark matter.  Modeling such dark matter systems has been a major focus
of astrophysicists, with much of that effort directed at computational
techniques.  Not surprisingly, a comprehensive understanding of the
evolution of these self-gravitating systems still eludes us,  since it
involves the collective nonlinear dynamics of many-particle systems
interacting via long-range forces described by the Vlasov equation.
As a step towards developing a clearer picture of collisionless
self-gravitating relaxation, we analyze the linearized dynamics of
isolated one-dimensional systems near thermal equilibrium by expanding
their phase space distribution functions $f(x,v)$ in terms of Hermite
functions in the velocity variable, and Legendre functions involving
the position variable.  This approach produces a picture of
phase-space evolution in terms of expansion coefficients, rather than
spatial and velocity variables.  We obtain equations of motion for the
expansion coefficients for both test-particle distributions and
self-gravitating linear perturbations of thermal equilibrium.
$N$-body simulations of perturbed equilibria are performed and found
to be in excellent agreement with the expansion coefficient approach
over a time duration that depends on the size of the expansion series
used.

\end{abstract}

\begin{keywords}
galaxies:kinematics and dynamics -- dark matter.
\end{keywords}

\section{Introduction}\label{intro}

Over the past several decades, much evidence has been compiled
supporting the idea that the baryonic mass visible in galaxies (stars,
gas, and dust) comprises a small fraction of the total gravitating
mass of such a system.  The earliest evidence comes from observations
of galactic motions within larger galaxy cluster systems.  Individual
galaxies had velocities that were too large to remain bound to the
cluster, given the inferred amount of stellar mass \citep{z37}. However,
the uncertainties associated with this analysis were large, and it
took several more decades for more conclusive evidence to emerge.  The
rotation curves (circular speed versus galactocentric distance) of
spiral galaxies are considered to be one of the clearest pieces of
evidence for what has become known as dark matter surrounding
galaxies.  In general, these curves show circular speeds of stars and
gas in spiral galaxies following solid-body-like rotation near their
centers, then reaching a nearly constant value \citep{rf70}.  This
contrasts with predictions based on the observed stellar/gas mass
distributions in these galaxies, where the circular speed should peak
and then decrease in the outer regions of a galaxy.  Further studies
of stellar kinematics in elliptical galaxies can hint at the need for
dark matter, but the dynamics of such systems are more complex than
for spiral galaxies, and interpretations are not as clear \citep{r03}.

In parallel with these inferences from galaxy dynamics, the idea of
dark matter has also been supported by cosmological investigations.
Numerical simulations of large-scale structure formation in the
universe can reproduce the observed filamentary structure of galaxy
clusters if dark matter is included \citep{nfw96,s05}.  Observations of
the cosmic microwave background reveal features that can be described
best when roughly 25\% of the mass in the universe is dark
matter \citep{wmap}.  A third route of evidence for dark matter around
galaxies involves observations of gravitational lensing.  Locations
and magnifications of images of distant galaxies and quasars that form
when their light is bent around intervening galaxies (or clusters of
galaxies) indicate that the lensing galaxies should have masses larger
than what can be accounted for from their visible
components \citep{lens,ws11}.

The current paradigm assumes that dark matter must act
collisionlessly.  The argument supporting this assumption is as
follows.  Observations indicating the presence of dark matter have not
shown indications of an edge to the dark matter halo.  For example,
there are no isolated spiral galaxy rotation curves where the circular
speed of gas begins to show a Keplerian decrease at some distance.  As
a result, it is assumed that the dark matter structures around
galaxies have much larger spatial extent than the visible components.
The baryons that will eventually form stars (mostly Hydrogen gas) are
initially mixed with the dark matter over these larger volumes, but
the baryons will self-interact via forces other than gravity.  This
gives the baryons a cooling mechanism that is unavailable to
dissipationless dark matter and allows gas to radiate energy away and
sink towards the center of the dark matter structure (typically
referred to as a halo).  Further, collisional effects would lead to
halos with more spherical shapes that observations of galaxy clusters
would allow \citep{mefg95}.

The previously mentioned cosmological simulations of structure
formation have done more than simply suggest the reality of dark
matter, they have also predicted its behavior on the scale of
galaxies.  It is generally agreed upon in the simulation community
that dark matter halo mass density profiles have central cusps $\rho
\propto r^{-\gamma}$ where $\gamma \approx 1$.  The logarithmic
density profiles then monotonically steepen as one moves away from the
center \citep[\eg][]{nfw97,n04}.  The consistency of the density
behavior across mass scales, initial conditions, and simulation
methods suggests that some simple underlying physics is at play in
these self-gravitating collisionless systems.  Further investigations
into the kinematics of dark matter systems have likewise heightened
the suggestion of a fundamental physical process driving the formation
of mechanical equilibrium dark matter halos \citep{tn01,hm06,ld11}.
Investigations of these three-dimensional (3-d) systems involve a wide
range of modes of evolution that contribute to the relaxation from
initial conditions to a final equilibrium state.  The radial orbit
instability \citep{ma85}, along with evaporation and ejection
\citep[][Chapter 7]{bt87}, are examples of these modes.  

The dynamics of collisionless systems of particles interacting
via long-range forces is studied with the Vlasov and Poisson
equations, which describe the evolution of the distribution of
particles in phase space.  There is a rich and on-going record of
previous work related to this topic, both in astronomy and plasma
physics.  A good introduction to the literature related to
Vlasov-solver techniques may be found in \citet{ac05}, where those
authors categorize previous methods of solution.  One category is the
``water bag'' solution, first discussed by \citet{dp62} and utilized
in an astrophysical setting by \citet{hf67}.  This method tracks the
boundary of a patch of a region of phase-space, inside which the
distribution function is constant, but as the occupied phase-space
becomes more filamentary due to phase mixing, following the boundary
becomes more computationally costly.  Another route to solutions
relies on grid-based techniques.  An early, popular implementation
came from plasma physics \citep{ck76}.  Astrophysical situations
following the lead of \citet{ck76} arrived slightly later
\citep{f81,netal81,wetal81}.  Resolution effects negatively affect the
ability of grid-based codes to follow filamentary phase-space
structures, but there are routes available to minimize this drawback
\citep[\eg][]{hb13}.  Especially in higher dimensions, grid-based
codes can have resolution limited by computational constraints, but
the advent of graphics processing units is beginning to loosen these
bounds \citep{rf13}.  \citet{ac05} discuss an approach that is
reminiscent of smoothed particle hydrodynamics for following
phase-space evolution.  This technique has several attractive aspects,
but must, by its nature, deal with a coarse-grained distribution
function.

In this paper we will consider the evolution of a one-dimensional (1-d)
self-gravitating collisionless system \citep[which can also be
formulated as a ``sheet'' model, \eg][]{c50}.  Compared to 3-d models,
the 1-d model is easier to analyze while possessing the essential
features of 3-d systems --- attractive long range forces and
collisonless collective dynamics.  However, it lacks some of the
features of 3-d systems like angular momentum and tidal forces.
Though the model is formulated in terms of continuous distribution
functions, it can also be considered as the $N \rightarrow \infty$
limit of system of $N$ particles with masses $m$, interacting via
two-body gravitational attraction.  The evolution of the phase-space
distribution function is described by the the Vlasov equation (or
collisionless Boltzmann equation),
\begin{equation}\label{cbe}
\frac{\dif f}{\dif t} = \frac{\partial f}{\partial t} +
v\frac{\partial f}{\partial x} + a(x) \frac{\partial f}{\partial v}=0,
\end{equation}
where  $f(x,v;t)$ is the normalized distribution function
\begin{displaymath}
\int^{\infty}_{-\infty} \int^{\infty}_{-\infty} f(x,v;t) \, \dif x
\,\dif v = 1.
\end{displaymath}
The $t$ argument is implied in what follows. The density is obtained
simply by integrating over velocities
\begin{equation}\label{rhodef}
\lambda(x) =  M \int^{\infty}_{-\infty} f(x,v)\, \dif v ,
\end{equation}
where $M$ is the total mass of the system (mass per unit area in the
sheet model, $Nm$ for particles).  The acceleration $a(x)$ for 1-d
systems is calculated  by simply taking the difference of the total
masses on each side of $x$,
\begin{eqnarray}\label{adef}
a(x) &= &-g \int^{x}_{-\infty}\lambda(s) \dif s +g
\int^{\infty}_{x}\lambda(s) \dif s \\ &=&g(M_>-M_<)\nonumber.
\end{eqnarray}
Note the long-range nature of the interaction, which couples particles
through the distance between them.  Likewise, the density is non-local
in phase space, in that it involves an integral over velocities.

Studies of such 1-d systems have a long history \citep{c50}.  In
general, much of the work can be categorized by dealing with either
cosmological conditions \citep[\eg][]{yg98,v06,mrl07,bjs13} where an
additional non-self-gravitating potential energy term is included in
the Hamiltonian (or periodic boundary conditions are used), or
isolated systems where self-gravity is the only source of potential
\citep[\eg][]{rm88,kk01,s13}.  Within each of these categories, a
variety of initial conditions have been investigated.  Broadly
speaking, initial conditions are typically near-equilibrium
\citep[\eg][]{rm87} or far-from-equilibrium \citep[\eg][]{jw11}.
Several investigations \citep[\eg][]{k77,m90,w91,boy11} have focused
on solving linearized versions of Equation~\ref{cbe}, a topic we
discuss in more detail in \S~\ref{linear}.  The situations we
investigate are isolated systems near thermal equilibrium.  Such
non-equilibrium systems might be considered to be in the final stages
of condensation from uniform cosmological conditions, or perhaps in
the aftermath of a collision in which two systems coalesce/pass
through one another.  We note that the absence of tidal forces in 1-d
guarantees that non-overlapping systems can be considered as isolated,
so that our discussion also applies to clusters of non-overlapping
systems between encounters. 

What follows here is a discussion of a method for finding solutions to
a linearized version of Equation~\ref{cbe}.  Our solution to the
linearized Vlasov equation is different than many previous solutions
that use an action-angle approach \citep[\eg][]{k77,m90,w91,boy11}.
In that approach, the linearized Vlasov equation is transformed from a
description in position-velocity coordinates to an action-angle
representation.  The resulting partial differential equation can then
be reduced to an algebraic equation using Fourier and Laplace
transforms.  The evolution of a perturbed distribution function in the
new variables is then, in principle, determined.  With appropriate
transforms, potential-density pairs can then be found.  A major
strength of this approach lies in its general nature; perturbations
are taken about any equilibrium state and external forces are
incorporated simply.  Another advantage taken by the action-angle
approach is the simplification afforded by the Laplace transform to
remove the time-derivative term in the Vlasov equation.  But while the
various transforms can simplify the differential equation, they also
introduce the need for inverse transforms.  Another issue arises from
the connections between the position-velocity and action-angle
representations.  If one wants to define the initial system in terms
of position-velocity coordinates, then transformations such as
$x(E,\theta)$ and $v(E,\theta)$ must be known (here, $E$ is energy and
$\theta$ is the conjugate angle coordinate).  Likewise, if the final
position-velocity distribution function is desired, one needs
$\theta(x,v)$.  In general, the tranformations between variables
involve numerical integrals, with continuous parameters, which have to
be inverted.  Finally, density and potential functions are not simply
connected to action-angle solutions and require expansions in
bi-orthonormal functions.

Our approach is to expand the distribution function in terms of
orthogonal functions. This method has been used previously, with
Hermite polynomials to describe the velocity aspect of distributions
\citep{rm88}, and Fourier expansions for the position for cosmological
models with periodic boundary conditions \citep{am09,rm87}.  The form
of the thermal equilibrium distribution function for isolated systems
very naturally suggests the use of Hermite polynomials for the
velocity and Legendre polynomials in $\tanh(x)$ for the position.  The
resulting linear set of equations of motion link the expansion
coefficients $c_{m,n}(t)$.  In this notation, $m$ and $n$ are the
orders of the Hermite and Legendre polynomials, respectively.  There
are few couplings between the coefficients --- in fact, the couplings
are local, in that they are only between neighbors on the $(m,n)$
grid.  This is rather fortuitous in light of the long-range nature of
the forces, and gives a simple local continuity-type evolution of
coefficients on the $(m,n)$ grid.  Furthermore, the method provides an
alternative to $N$-body simulations that yields smooth distribution
functions, at least for modestly perturbed systems.
 
The outline of this paper is as follows. In section~\ref{thermeq}, the
properties of thermal equilibrium are summarized.  The expansion of
the distribution function in terms of Hermite-Legendre polynomials is
developed in \S~\ref{ortho}.  The Vlasov equation is linearized and
the equations of motion of the expansion coefficients are obtained in
Section~\ref{linear}.  The behavior of solutions is discussed in
Section~\ref{sols}.  Numerical considerations and comparisons with
$N$-body simulations are presented in \S~\ref{sims}.  Conclusions and
future directions are discussed in Section~\ref{discuss}. 
 
\section{Thermal Equilibrium}\label{thermeq}

Based on the structure of Equation~\ref{cbe}, any function of the
single particle energy,
\begin{equation}\label{spece}
\epsilon = \frac{1}{2}mv^2 + m\phi(x),
\end{equation}
is a time-independent solution.   Thermal equilibrium is a special
case which case the distribution function has the separable Boltzmann
form
\begin{equation}\label{boltz}
f_0(\epsilon) = A e^{-\beta \epsilon}=Ae^{-\frac{\beta m v^2}{2}}
e^{-\beta m \phi},
\end{equation}
where $\beta\equiv 1/k_BT=1/\left< {mv^2}\right>$ is an energy scale
(commonly referred to as the inverse temperature),  and $A$ is a
normalization constant.  Upon substitution of
Equations~\ref{spece}-\ref{boltz} into Equation~\ref{cbe}, it is
straightforward to obtain the thermal equilibrium distribution
function, which is commonly written as,
\begin{equation}\label{ftherm0}
f_0(x,v) = A \sech^2{(\frac{\beta gm \mt}{2} x)} e^{-\frac{\beta
mv^2}{2}},
\end{equation}
where $A =(g\mt/4)\sqrt{\beta^3m^3/2\pi}$.  This is the
well-known result that is presented in \citet{c50}.  \citet{r71} has
shown that this is also the $N \rightarrow \infty$ limit for a system
of $N$ equal mass particles in both canonical and micro-canonical
ensembles.

The potential corresponding to this equilibrium is,
\begin{eqnarray}
\phi_0(x) &=&  \int^{\infty}_{-\infty}g|{x-s}|\lambda(s) \dif s
\nonumber\\ &=&\frac{2}{\beta m}\ln{(2 \cosh{\frac{\beta g m \mt}{2}
x})},
\end{eqnarray}
from which we obtain the acceleration
\begin{equation}
a_0(x)=-\frac{\partial \phi_0(x)}{\partial x}=-g \mt\tanh \frac{\beta
g m \mt}{2} x.
\end{equation}
In terms of the quantities defined, the kinetic energy of the
equilibrium state is
\begin{equation}
K_0=\mt\int_{-\infty}^{\infty}\frac{v^2}{2} f_0(x,v) \,{\dif x}\,{\dif
v}=\frac{N}{2\beta}.
\end{equation}
The equilibrium potential energy is likewise given by
\begin{equation}
U_0=\frac{1}{2}\int_{-\infty}^{\infty}\lambda_0(x) \phi_0(x,v) \,{\dif
x}\,{\dif v}=\frac{N}{\beta} =2K_0,
\end{equation}
as required by the virial theorem for one dimension.

The Boltzmann nature of the one-dimensional self-gravitating system is
a vital difference from the three-dimensional case.  Mechanical
equilibria of realistic three-dimensional self-gravitating systems
always contain gradients in the kinetic temperature, $T_{\rm K}
\propto \langle v^2 \rangle$, that act as pressure support against
gravity.  Only the infinite mass and energy isothermal sphere has a
constant temperature.  This one-dimensional distribution function is a
true thermal equilibrium, as the kinetic temperature is uniform
throughout the equilibrium system.

For simplicity, we transform to dimensionless coordinates using the
definitions,
\begin{displaymath}
\chi = \frac{\beta g m \mt}{2} x \quad\mbox{and} \quad
\varpi = \sqrt{\frac{\beta m}{2}} v.
\end{displaymath}
 The scaled equilibrium distribution function is,
\begin{equation}
\fst_0(\chi,\varpi) = \frac{2}{\beta g m \mt }\sqrt{\frac{2}{\beta m}}
f_0 = \frac{1}{2\sqrt{\pi}} \sech^2{\chi} \, e^{-\varpi^2}.
\end{equation}
where tildes are used to indicate dimensionless functions, when a
distinction is necessary.  The Vlasov equation transforms to,
\begin{equation}\label{cbe2}
\frac{\partial \fst}{\partial \tau} +
\varpi\frac{\partial \fst}{\partial \chi} + \alpha(\chi) 
\frac{\partial \fst}{\partial \varpi}=0,
\end{equation}
where $\tau=\sqrt{\beta m/2} \,g \mt t$ is the dimensionless
time and $\alpha(\chi)=a/(g \mt)$ is the dimensionless
acceleration function.

\section{Orthogonal Polynomials}\label{ortho}

The form of the equilibrium distribution function suggests a set of
orthogonal functions to use as a basis for a polynomial expansion.
We consider the expansion,
\begin{equation}\label{fsum}
\fst(\chi,\varpi) = \sum_{i,j} c_{i,j}
G_{ij}(\chi,\varpi) \fst_0(\chi,\varpi),
\end{equation}
where the $c_{i,j}$ are real expansion coefficients.  The $G_{ij}$ are
functions defined by
\begin{equation}\label{Gdef}
G_{ij}(\chi,\varpi) =\sqrt{\frac{2j+1}{2^{i} i!}} 
H_i(\varpi) P_j(\tanh \chi),
\end{equation}
where the $H_i$ are Hermite polynomials of order $i$, and the $P_j$
are Legendre polynomials of order $j$. The $G_{ij}$ are constructed to
be orthonormal to $f_0$, which serves as a weighting function,
\begin{equation}
\int_{-\infty}^{\infty}
\int_{-\infty}^{\infty}G_{ij}(\chi,\varpi)G_{i^{\prime}
j^{\prime}}(\chi,\varpi)(\varpi) \fst_0(\chi,\varpi) \, \dif \chi\,
\dif \varpi = \delta_{i i^{\prime}} \delta_{j j^{\prime}}.
\end{equation}

We routinely use the Hermite polynomial orthogonality condition,
\begin{equation}\label{Hortho}
\int_{-\infty}^{\infty} H_i(\varpi) H_{i^{ \prime}}(\varpi)
e^{-\varpi^2} \, \dif \varpi = \delta_{i i^{\prime}} 2^i \sqrt{\pi}
i!,
\end{equation}
where $\delta$ is the Kronecker delta.  For the Legendre orthogonality
condition, we can eliminate the factor $\sech^2\chi$ with the change
of variables $u=\tanh{\chi}$ and $\dif u = \sech^2{\chi} \dif \chi$, 
\begin{eqnarray}\label{Portho}
\int_{-\infty}^{\infty} P_j(\tanh\chi) P_{j^{
\prime}}(\tanh\chi)\sech^2\chi \, \dif \chi & = & \nonumber \\
\int_{-1}^{1} P_j(u) P_{j^{\prime}}(u) \, \dif u & = & 
\delta_{j j\prime} \frac{2}{2j+1}.
\end{eqnarray}
Note that this substitution also maps infinite limits on any $\chi$
integral to the interval [-1,1].

At thermal equilibrium, only the $i=0$, $j=0$ coefficient is nonzero.
For an arbitrary distribution function $\fst (\chi, \varpi)$ perturbed
from thermal equilibrium, the coefficients can be determined from
\begin{equation}\label{coeffs}
c_{i,j} = \int_{-\infty}^{\infty} \int_{-\infty}^{\infty}
G_{i,j} (\chi, \varpi)\fst (\chi, \varpi) \, \dif\chi  \, \dif\varpi .
\end{equation}
This equation represents a transformation from phase space to a
discrete $(i,j)$ grid of coefficients.

The expansion dictates that all mass must derive from the (0,0) term,
\begin{eqnarray*}
{\tilde M}_{i,j} \equiv \frac{M_{i,j}}{M}& = &
\int_{-\infty}^{\infty} \int_{-\infty}^{\infty} \fst_{i,j} \, \dif
\chi \, \dif \varpi \nonumber \\ & = &   \int_{-\infty}^{\infty}
\int_{-\infty}^{\infty}  c_{i,j} G_{ij}(\chi,\varpi)
\fst_0(\chi,\varpi)\, \dif \chi \, \dif \varpi \nonumber \\ & = &
c_{i,j}  \delta_{i0} \delta_{j0},
\end{eqnarray*}
from which we obtain  $\tilde{M}_{0,0}=c_{0,0}=1$.  In a similar
fashion, one can see that mass density $\rhot (\chi)$ derives only
from $i=0$ terms,
\begin{eqnarray}\label{lindens}
\rhot(\chi) & = &\int_{-\infty}^{\infty} \fst_{i,j} \,  \dif \varpi
\nonumber \\ & = &  \int_{-\infty}^{\infty} \sum_{i,j}  c_{i,j}
G_{ij}(\chi,\varpi) \fst_0(\chi,\varpi) \, \dif \varpi \nonumber \\
& = & \sum_j c_{0,j} \sqrt{2j+1}P_j(\tanh\chi)\rhot_0(\chi).
\end{eqnarray}

\section{Linear Perturbations}\label{linear}

We now use the orthogonal polynomials developed in the previous
section as a basis to study the dynamics of perturbations from thermal
equilibrium.  We consider distribution functions of the form,
\begin{equation}
\fst=\fst_0 + \delta \fst_1,
\end{equation}
where $\fst_1$ is the perturbing function and $\delta \ll 1$ is an
expansion parameter.

Using this perturbed $\fst$ in Equation~\ref{cbe} produces a modified
Vlasov equation for the perturbing function (in terms of the
previously defined dimensionless quantities),
\begin{equation}\label{pcbe}
\frac{\partial \fst_1}{\partial \tau} + \varpi \frac{\partial
\fst_1}{\partial \chi} + \alpha_0(\chi) \frac{\partial
\fst_1}{\partial \varpi} =2\varpi \alpha_1(\chi)  \fst_0
\end{equation}
where we have used $\partial \fst_0/\partial \varpi = -2\varpi
\fst_0$.  The accelerations are given by
\begin{displaymath}
\alpha_0(\chi)=-\int_{-\infty}^{\chi}\rhot_0(\chi^\prime)\,
\dif\chi^\prime +\int_{\chi}^{\infty}\rhot_0(\chi^\prime)\,
\dif\chi^\prime = -\tanh \chi,
\end{displaymath}
\begin{equation}\label{alpha1}
\alpha_1(\chi)=-\int_{-\infty}^{\chi}\rhot_1(\chi^\prime)\,
\dif\chi^\prime +\int_{\chi}^{\infty}\rhot_1(\chi^\prime)\,
\dif\chi^\prime.
\end{equation}
The term $ 2\varpi \alpha_1(\chi)  \fst_0$ is required by Newton's
Third Law.  In this equation, it has been written on the right hand
side to signify that it is neither a convective nor an advective term.
In fact, the right hand side is best characterized as a collision term
as it represents the deflection of particles into and out of
equilibrium due to the perturbation.  Here, we ignore the second-order
term describing the self-interaction of the perturbing particles,
$\alpha_1\partial \fst_1 /\partial \varpi$.

We now express the perturbing distribution function in terms of the
Hermite and Legendre polynomials discussed earlier,
\begin{equation}\label{f1expansion}
\fst_1=\sum_{i,j}  c_{i,j} \sqrt{\frac{2j+1}{2^i\,i!}}H_i(\varpi)
P_j(\tanh \chi) \frac{\sech^2 \chi e^{-\varpi^2}}{2\sqrt{\pi}},
\end{equation}
where $c_{0,0}=0$ since the equilibrium contribution has already been
removed.  This guarantees that the perturbations are massless.  Using
Equation~\ref{lindens} in Equation~\ref{alpha1}, the perturbing
acceleration is found to be,
\begin{eqnarray}\label{alpha1b}
\alpha_1(\chi) & = & \sum_j c_{0,j}\sqrt{2j+1}\left[ \int_{\chi}^{\infty}
P_j (\tanh \chi^{\prime}) \frac{\sech^2}{2} \chi^{\prime} \,
\dif \chi^{\prime} - \right. \nonumber \\
 & & \left. \int_{-\infty}^{\chi} P_j(\tanh
\chi^{\prime}) \frac{\sech^2}{2} \chi^{\prime} \, \dif
\chi^{\prime} \right].
\end{eqnarray}
Upon making the  substitution $u=\tanh \chi$, $\sech^2 \chi = 1-u^2$,
$\dif u = (1-u^2)\, \dif \chi$, this simplifies to,
\begin{equation}
\alpha_1(u)  = \frac{1}{2}\left[ \sum_j c_{0,j}\sqrt{2j+1} \left(
\int_{u}^{1} P_j (u^{\prime}) \, \dif u^{\prime} - \int_{-1}^{u} P_j
(u^{\prime}) \, \dif u^{\prime} \right) \right] .
\end{equation}
In terms of the $u$ variable, the modified Vlasov equation
(Equation~\ref{pcbe}) becomes,
\begin{equation}\label{pcbe3}
\frac{\partial \fst_1}{\partial \tau} + \varpi (1-u^2) \frac{\partial
\fst_1}{\partial u} - u \frac{\partial \fst_1}{\partial
\varpi} -2 \varpi \alpha_1(u)\fst_0=0.
\end{equation}
Substituting Equation~\ref{f1expansion} and canceling a common factor
of $\fst_0$ produces,
\begin{eqnarray}\label{boltzpen}
\lefteqn{\sum_{i,j}\left\{ \dot{c}_{i,j} H_i P_j + c_{i,j}  \varpi H_i
(1-u^2)\frac{\partial P_j}{\partial u} -c_{i,j}  u P_j \frac{\partial
H_i}{\partial \varpi}\right\} -} & & \nonumber \\
& & 2 \varpi \alpha_1(u)=0.
\end{eqnarray}

We make use of the following recursion relations to obtain equations
of motion for the $c_{i,j}$; 
\begin{equation}
(1-u^2) \frac{\partial P_j(u)}{\partial u} = \frac{j(j+1)}{2j+1}\left[
P_{j-1}(u) + P_{j+1}(u) \right],
\end{equation}
\begin{equation}
\frac{\partial H_i(\varpi)}{\partial \varpi} = 2i H_{i-1}(\varpi),
\end{equation}
\begin{equation}
uP_j(u) = \frac{1}{2j+1} \left[ (j+1)P_{j+1}(u) + jP_{j-1}(u) \right],
\end{equation}
\begin{equation}
2\varpi H_i(\varpi) = H_{i+1}(\varpi) + 2iH_{i-1}(\varpi),
\end{equation}
and
\begin{equation}
\int P_j(u) \, \dif u = \frac{P_{j+1}(u) - P_{j-1}(u)}{2j+1}.\label{recurint}
\end{equation}
Substituting these relations into Equation~\ref{boltzpen}, using
the fact that $P_n(1)=1$ and $P_n(-1)=(-1)^n$ in the simplification
of $\alpha_1$, results in,
\begin{eqnarray}
 & & \sum_{i,j}\sqrt{\frac{2j+1}{2^i\,i!}} \left\{\dot{c}_{i,j} H_i
 P_j + c_{i,j} \left[ \frac{j(j+1)}{2(2j+1)} H_{i+1} P_{j-1} - \right.
\right. \nonumber \\
 & & \left. \left. \frac{j(j+1)}{2(2j+1)} H_{i+1} P_{j+1} + 
\frac{ij(j-1)}{2j+1}
 H_{i-1,j-1} - \right. \right. \\
 & & \left. \left. \frac{i(j+1)(j+2)}{2j+1} H_{i-1} P_{j+1} +
 \delta_{0,i} \frac{1}{2j+1} H_1 \left[ P_{j+1} - P_{j-1}
 \right] \right]\right\}=0 \nonumber.
\end{eqnarray}
The term containing the Kroenecker delta corresponds to the $ 2\varpi
\alpha_1(\chi) \fst_0 $ term in Equation~\ref{pcbe} and is zero except
when $i=0$.  Finally, we obtain the equations of motion for the
coefficients by multiplying this expression by $G_{m,n} \fst_0$,
integrating over $\varpi$ and $u$, and making use of the orthogonality
relations Equations~\ref{Hortho}-\ref{Portho}.  The resulting
expressions have the form,
\begin{eqnarray}\label{cbelp}
\dot{c}_{m,n} & = & L_{m,n}^{m-1,n-1} \, c_{m-1,n-1} +
  L_{m,n}^{m-1,n+1} \, c_{m-1,n+1} +\nonumber\\ & & L_{m,n}^{m+1,n-1}
  \, c_{m+1,n-1}  + L_{m,n}^{m+1,n+1} \, c_{m+1,n+1} ,
\end{eqnarray}
where the matrix elements $L_{m,n}^{i,j}$ are given by 
\begin{eqnarray}\label{Ldef}
L_{m,n}^{m-1,n-1}&=&\frac{\sqrt{m}(n-1)n-2\delta_{1,m}}
{\sqrt{2(2n+1)(2n-1)}},\nonumber \\
L_{m,n}^{m-1,n+1}&=&-\frac{\sqrt{m}(n+2)(n+1)-2\delta_{1,m}}
{\sqrt{2(2n+1)(2n+3)}},\nonumber \\
L_{m,n}^{m+1,n-1}&=&\frac{\sqrt{m+1}(n+1)n}
{\sqrt{2(2n+1)(2n-1)}},\nonumber \\ 
L_{m,n}^{m+1,n+1}&=&- \frac{\sqrt{m+1}(n+1)n}{\sqrt{2(2n+1)(2n+3)}},
\end{eqnarray}
where $m,n,i,j \ge 0$. The test-particle case is obtained by omitting
the Kronecker $\delta_{1,m}$ terms. 

Equations~\ref{cbelp}-\ref{Ldef} are the main results of this paper.
For {\it linearized} dynamics the $c_{m,n}$ evolve by coupling to
diagonal neighbors only. This is somewhat surprising in light of the
long-range nature of the forces, and can be traced back to the
recursion relation Equation~\ref{recurint}, that replaces the integral over
$\chi$ in the calculation of $\alpha_1$. Because of this
nearest-diagonal-neighbor coupling, the even parity and odd parity
modes completely decouple, where the parity is given by $(-1)^{m+n}$.
For simplicity, we shall concern ourselves with the even parity modes
only, and set all the odd parity coefficients to zero. This
automatically guarantees that the center of mass velocity and position
are zero,  $\left< \varpi \right>=\left< \chi \right>=0$.

As discussed in Section~\ref{intro}, an action-angle approach is
often used to solve the linearized Vlasov equation.  In contrast to
this approach, our solution is specific to thermal equilibrium in the
absence of any external potential, a comparative weakness.
Additionally, our approach does not take advantage of a Laplace
transform to remove the time derivative in the Vlasov equation,
leading to the need to solve an ordinary differential equation.  Such
a transform might provide additional insight into our method, and we
leave that as a future direction of work.  Despite these facts, we
suggest that our solution leads to a more straightforward
interpretation of perturbation evolution for self-gravitating systems
than the action-angle approach affords.  The more specific link
between equilibrium and our choice of transforms leads to the
simplifications of the nearest-neighbor and decoupling behaviors
discussed in the previous paragraph and provides an easily visualized
evolution in ($m,n$) space.  Additionally, our method makes the
initial value problem simple -- specific $(x,v)$ perturbations can be
easily represented by a set of $c_{m,n}$ via a simple projection
operation.  Our approach also provides a straightforward calculation
of the spatial dependence of acceleration and density functions.

\section{Behavior of Solutions}\label{sols}

Although we do not attempt to find a general solution of
Equation~\ref{cbelp} on the entire $(m,n)$ domain, we can sketch the
behavior of coefficients by considering some restricted situations.
As a first case, imagine that only the $c_{0,2}$, $c_{1,1}$,
$c_{2,0}$, and $c_{2,2}$ coefficients are available to be non-zero.
This truncated $2\times 2$ system, although unrealistic, should be
describe the behavior of $c_{1,1}$  for short times.   The equations
of motion found from Equations~\ref{cbelp}-\ref{Ldef},
\begin{eqnarray}\label{sho}
\dot{c}_{1,1} & = & -\frac{2\sqrt{2}}{\sqrt{30}} c_{2,2} +
\frac{2\sqrt{10}}{\sqrt{30}} c_{2,0} - \left(
\frac{6}{\sqrt{30}}-\frac{2}{\sqrt{30}}\right) c_{0,2}, \nonumber \\
\dot{c}_{0,2} & = & \frac{6}{\sqrt{30}} c_{1,1}, \nonumber \\ 
\dot{c}_{2,0} & = & -\frac{2\sqrt{10}}{\sqrt{30}} c_{1,1},\nonumber \\
\dot{c}_{2,2} & = & \frac{2\sqrt{2}}{\sqrt{30}} c_{1,1}.
\end{eqnarray}
The term in parentheses has the terms due to $\alpha_0$ and $\alpha_1$
separated; for test particles only the first term is present. Taking
another time derivative and substituting gives $\ddot{c}_{1,1} =
-(12/5) c_{1,1}$ for self-gravitation [$\ddot{c}_{1,1} = -(14/5)
c_{1,1}$ for test particles], indicating that the $c_{1,1}$
coefficient value should oscillate with a period
$T_{self}=2\pi/\sqrt{12/5}\approx 4$ for self-gravitation
($T_{test}\approx3.75$ for test particles).   The other coefficients
$c_{0,2}$, $c_{2,0}$ and  $c_{2,2}$ are proportional to ${\dot
c}_{1,1}$ so they oscillate $90^\circ$ out of phase with $c_{1,1}$.
In phase space, this behavior is a simple oscillation of two density
peaks back and forth through the center of mass.

Results of solving Equation~\ref{cbelp} and numerical simulations (see
\S~\ref{sims}) agree with the frequencies found here for early times.
Time-independent solutions exist as well, even for this simple system.
For example,
\begin{displaymath}
c_{1,1} = 0,\,c_{2,0}=\sqrt{2/7}C,\,c_{0,2}=\sqrt{5/7}C,\,c_{2,2}=0,
\end{displaymath}
is a normalized solution, where $C$ is a constant. In these respects,
the self-gravitating system is similar to a distribution of harmonic
oscillators. However, the harmonic oscillator coefficients change only
by coupling between coefficients with the same $m+n$ values; for
example, $\dot{c}_{1,1}^{\rm harmonic} = 2 c_{2,0}^{\rm harmonic} - 2
c_{0,2}^{\rm harmonic}$.  The difference between this expression and
the analogous relation in Equation~\ref{sho} reflects the fact that
the particles in the harmonic potential do not experience phase
mixing, while those in the gravitational case do.

The most serious limitation with this simple picture is that we have
ignored the coupling to higher polynomial terms.  The result of this
coupling is most easily analyzed in the large $(m,n)$ limit with
gradual variations of $c_{m,n}$ on the $(m,n)$ grid.  Taking a second
derivative of Equation~\ref{cbelp} and substituting the appropriate
derivatives, one obtains, after specifying that $m \gg 1$ and $n \gg
1$, 
\begin{eqnarray}\label{tpshm2}
\ddot{c}_{m,n} & \approx & \frac{m n^2}{4} (c_{m+2,n+2} -
2c_{m+2,n} + c_{m+2,n-2}) + \nonumber \\
 & & \frac{mn^2}{2} (c_{m,n+2} - 2c_{m,n} + c_{m,n-2}) + \nonumber \\
 & & \frac{mn^2}{4} (c_{m-2,n+2} - 2c_{m-2,n} + c_{m-2,n-2}).
\end{eqnarray}
Each of the terms in parentheses in Equation~\ref{tpshm2} is a finite
difference approximation to a second derivative, if $n$ were
considered a continuous variable.  If we consider only slowly varying
dependence on $m$, then the second line of Equation~\ref{tpshm2} is
approximately equal to the sum of the first and third lines, so that
Equation~\ref{tpshm2} simplifies to,
\begin{equation}
\ddot{c}_{m,n} \approx (mn^2) \frac{\partial^2 c_{m,n}}{\partial n^2}.
\end{equation}
This wave equation form suggests a simple visualization where  the
distribution in $(m,n)$ is propagated toward higher $n$ (at increasing
speeds $\approx \sqrt{m}n$).  Combining the findings of these two
restricted cases, we develop a picture of how coefficients behave
during evolution.  We expect low-order coefficients to oscillate but
with a decreasing amplitude as they effectively radiate to higher-order
coefficients. This contrasts with the phase-space description where
the distribution function becomes filamentary as it ``winds up'' due to
dephasing. 

\section{Simulations}\label{sims}

In order to test the accuracy of the coefficient evolution approach in
describing systems near equilibrium, we have numerically integrated
Equation~\ref{cbelp} for simple initial perturbations. The results
have been compared with $N-$body simulations, for both test-particle
and self-gravitating dynamics.   Although the agreement is excellent
at short times and small perturbations, the differences are also
important as they give insights into the relative importance of
dephasing and self-gravitation in collisionless relaxation, finite $N$
and discreteness effects in $N-$body simulations, and the impact of
nonlinear effects. 

Numerical solutions to Equation~\ref{cbelp} have been found using a
midpoint method on a fixed-size $m_{\rm max} \times n_{\rm max}$ grid.
For initial conditions, we have used simple perturbations,  consisting
of one low-order excited mode, either the $c_{1,1}$, $c_{2,0}$, or
$c_{0,2}$.   For boundary conditions, we have tried several schemes,
but ultimately settled on fixed boundaries, \ie\ $c_{m,n_{\rm
max}}=c_{m_{\rm max},n}=0$.  This is the simplest choice that also
provides stable integrations.  For these boundary conditions, $m_{\rm
max}$ and $n_{\rm max}$ must be approximately 100 to get accurate
$d(t)$ values for low-order polynomials over their lifetime -
typically a few linear oscillation periods $\tau= 5T_{\rm self}
\approx 20$.  For stability, a fixed time step of
$d\tau=(10\sqrt{m_{\rm max}}n_{\rm max})^{-1}$ has been adopted for
the midpoint integration.  This ensures that each coefficient changes
by only a small relative amount during a time step.  A more detailed
discussion of boundary conditions is given in Section~\ref{discuss}.

For comparison with the coefficient evolution solutions,  $N$-body
simulations were performed of both test-particle and self-gravitating
systems.  For each case, 100 distinct realizations of a given initial
distribution function with $N$ particles have been evolved, forming
ensembles from which average quantities have been calculated.  We
typically adopt $N=1024$, but some ensembles with $N=2048$ have been
created to observe finite $N$ effects.  

Test-particle evolutions in the equilibrium potential utilize an
adaptive time step Runge-Kutta scheme to track particles.  Test-
particle accelerations are determined by the equilibrium
potential only.  Tolerances have been chosen so that the total
energies of test-particle systems experience fractional variations on
the order of $10^{-11}$ over the course of an evolution, providing
energy conservation comparable to the test-particle simulations.

The $N$-body simulations of self-gravitating systems have been carried
out using a heap-sort algorithm \citep{n03}, that takes advantage of
the 1-d system to achieve remarkable speed and accuracy (only
numerical round-off errors degrade the process).  Total system
energies fractionally vary by approximately $10^{-9}$ during
thousand-particle evolutions over $10^5 \, T_{\rm self}$, providing
energy conservation comparable to test-particle ensembles over tens of
oscillations. Typically $N=1024$ particles were used and averaged over
an ensemble of 100 random initial conditions.

\begin{figure}
\scalebox{0.65}{
\includegraphics{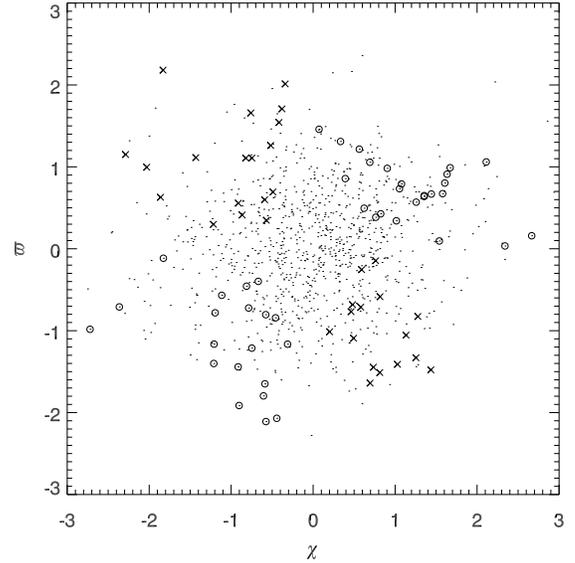}}
\caption{Initial snapshot of phase-space positions of an $N$-body
system experiencing a (1,1) perturbation of strength $b=0.1$.  The
circled particles are locations where the perturbed distribution
function is larger than the thermal equilibrium distribution, in
effect, the extra particles.  The crosses are particle locations that
are omitted because the perturbed distribution function is smaller
than the equilibrium distribution.}
\label{ip}
\end{figure}

Initial conditions with a single excited mode are generated by placing
particles at random phase-space coordinates according to the
probability distribution $\fst(\chi, \varpi)=\fst_0(\chi, \varpi)[1+b
G_{m,n}(\chi, \varpi)]$, where $b$ is the amplitude of the
perturbation.  Specifically, uniformly random coordinates are
generated in a region of phase space $-20 \le \chi,\varpi \le 20$.  A
third random variable $X \in [0,1]$ is then generated, and a particle
is placed at $(\chi,\varpi)$ if $X<\fst(\chi, \varpi)$.  In practice,
this process requires some caution since the perturbing terms can
result in negative distribution function values, at least for large
values of $\chi$ and $\varpi$.  In this case, the generated initial
distribution function would not be $\fst(\chi,\varpi)$ but ${\rm
max}[0,\fst(\chi,\varpi)]$, introducing spurious higher-order
polynomials.  Numerically, we have avoided this issue by setting
initial amplitudes of perturbing terms $c_{m,n}(t=0)=b$ small enough
to guarantee that the distribution is positive everywhere for $-20 \le
\chi,\varpi \le 20$.  As an example of how a perturbation impacts a
distribution function, Figure~\ref{ip} shows $N$-body initial
conditions for a $c_{1,1}(t=0)=0.1$ perturbation, which is about $1/3$
of the size where $\fst$ acquires negative regions.

The coefficients  $c_{m,n}$ can be calculated in the $N$-body
simulations using the discrete distribution function in terms of delta
functions,
\begin{equation}
\fst=\frac{1}{N}\sum_{i=1}^{N}\delta(\chi-\chi_i)\delta(\varpi-\varpi_i)
\end{equation}
where the $(\chi_i,\varpi_i)$ are the phase-space coordinates of the
$i$th particle. Substituting this into Equation~\ref{coeffs} gives 
\begin{eqnarray}\label{discretec}
c_{m,n}(\tau) &=& \int_{-\infty}^{\infty} \int_{-\infty}^{\infty}
G_{m,n} (\chi, \varpi)\fst (\chi, \varpi;\tau) \, \dif\chi  \,
\dif\varpi ,\nonumber \\ & = & \frac{1}{N}\sum_{i=1}^N G_{m,n}
(\chi_i(\tau), \varpi_i(\tau)). 
\end{eqnarray}
The ensemble averages of $N$-body simulations are denoted as $\left<
c_{m,n} \right>$ to distinguish them from coefficients evolved
according to Equation~\ref{cbelp}.  

\begin{figure}
\scalebox{0.5}{
\includegraphics{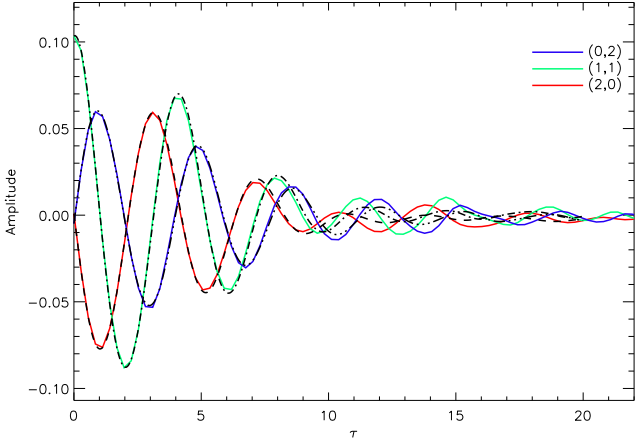}}
\caption{Test-particle coefficient evolutions for an initial
($m=1,n=1$) perturbation with strength $b=0.1$.  Time units are the
dimensionless time $\tau$ ($\tau \approx 4$ corresponds to one linear
oscillation period).  The solid lines show the behavior of ensemble
average coefficient values from simulations, while the thin broken
lines illustrate solutions to the coefficient equation of motion
(Equation~\ref{cbelp}, with Kronecker delta terms omitted in $L$).
For clarity, only the evolutions of the three lowest terms are
depicted.}
\label{compevtp}
\end{figure}

\begin{figure}
\scalebox{0.65}{
\includegraphics{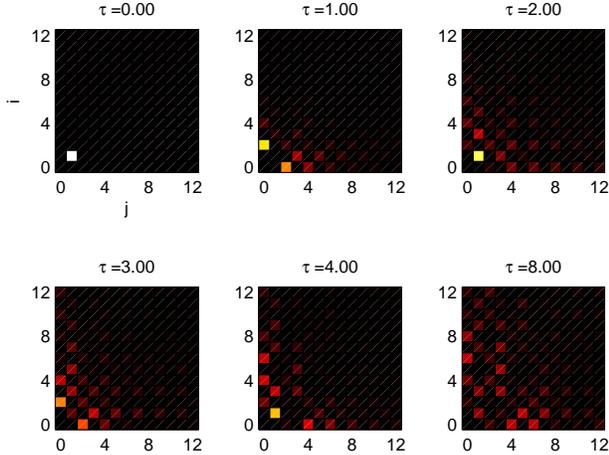}}
\caption{A time sequence of test-particle coefficient strengths
resulting from an initial ($m=1,n=1$) perturbation with strength
$b=0.1$.  Each frame is labeled with the dimensionless time unit
corresponding to that snapshot.  The shading of the squares represents
the relative values of $|c_{m,n}|$, with lighter shades for larger
amplitudes.  This sequence highlights the phase mixing behavior of the
coefficients, in particular, how amplitudes of lower-order terms
disperse to higher-order terms.}
\label{mnplane}
\end{figure}

\begin{figure}
\scalebox{0.5}{
\includegraphics{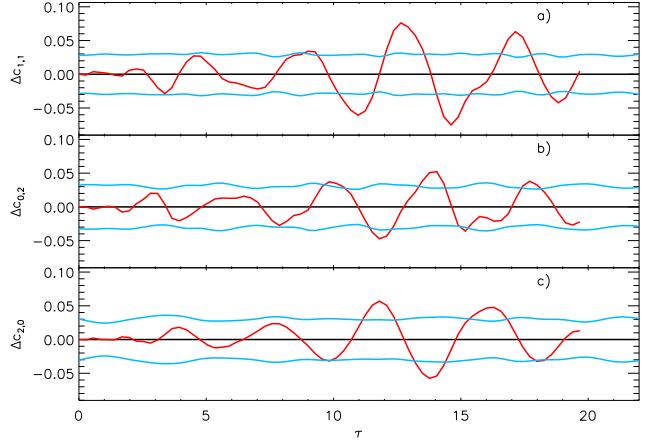}}
\caption{Differences between observed and predicted test-particle
coefficient evolutions for an initial ($m=1,n=1$) perturbation with
strength $b=0.1$.  Each panel shows how an ensemble-averaged
coefficient value from simulations $\left<c_{m,n}\right>$ compares to
the prediction from Equation~\ref{cbelp}: $c_{1,1}$ values are in
panel a, $c_{0,2}$ in panel b, $c_{2,0}$ in panel c.  Difference
values shown here are scaled by the initial value of the coefficient
of the perturbing term.  The horizontal lines represent the
``error in the mean'' range for ensemble averages.}
\label{diffevtp}
\end{figure}

Figure~\ref{compevtp} shows the time behavior  of the lowest-order
coefficients for the test-particle case, due to an initial (1,1)
perturbation, calculated by integrating Equation~\ref{cbelp}.  Also
shown are coefficients calculated from $N$-body simulations with test
particles. These lowest coefficients rapidly decrease, essentially
disappearing after only a few oscillations, as the amplitude of the
low-order term disperses between multiple higher-order terms.  This
would appear to be the result of phase mixing, where large scale
structures in phase-space (low order values) are transformed into ever
smaller scale features encoded in higher order polynomials.
Figure~\ref{mnplane} provides a clear view of how a coefficient's
amplitude disperses.  The initial (1,1) perturbation couples to
surrounding terms, next exciting (2,0) and (0,2) terms.  At later
times, the (1,1) term has a relatively small amplitude while numerous
higher-order terms have developed small, but non-zero, amplitudes.
The comparison between the ensemble averages $\left< c_{m,n} \right>$
and the coefficient $c_{m,n}$ is given as functions of time in
Figure~\ref{diffevtp}. The horizontal dashed lines indicate the sizes
of statistical errors in $\left< c_{m,n} \right>$ and provide a scale
for comparing differences. 

\begin{figure}
\scalebox{0.5}{
\includegraphics{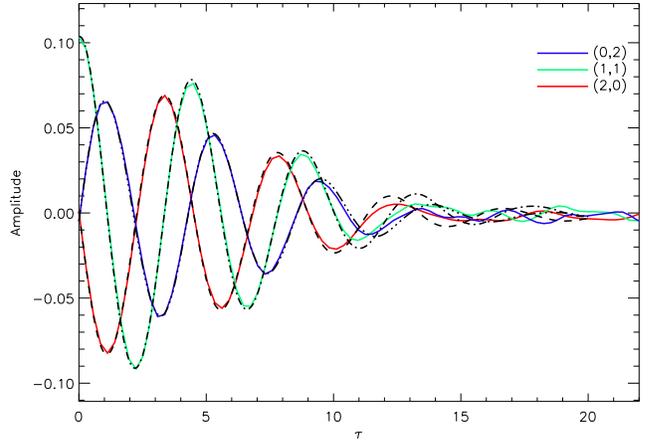}}
\caption{Coefficient evolutions for linear self-gravitation with an
initial (1,1) perturbation with strength $b=0.1$.  The solid lines
show the behavior of ensemble average coefficient values from
self-gravitating simulations, while the thin broken lines illustrate
solutions to the linear coefficient dynamics relationship
(Equation~\ref{cbelp}, with Kronecker delta terms included).  For
clarity, only the evolutions of the three lowest terms are depicted.}
\label{compevlp}
\end{figure}

\begin{figure}
\scalebox{0.5}{
\includegraphics{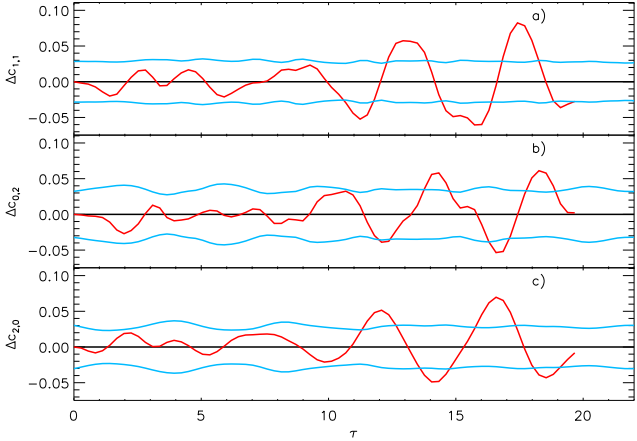}}
\caption{Differences between observed and predicted linear
perturbation coefficient evolutions for an initial (1,1) perturbation
with strength $b=0.1$.  Each panel shows how an ensemble-averaged
coefficient value from self-gravitating simulations
$\left<c_{m,n}\right>$ compares to the prediction from
Equation~\ref{cbelp}: $c_{1,1}$ values are in panel a, $c_{0,2}$ in
panel b, $c_{2,0}$ in panel c.  Difference values shown here are
scaled by the initial value of the coefficient of the perturbing term.
The horizontal lines represent the ``error in the mean'' range
for ensemble averages.}
\label{diffevlp}
\end{figure}

\begin{figure}
\scalebox{0.5}{
\includegraphics{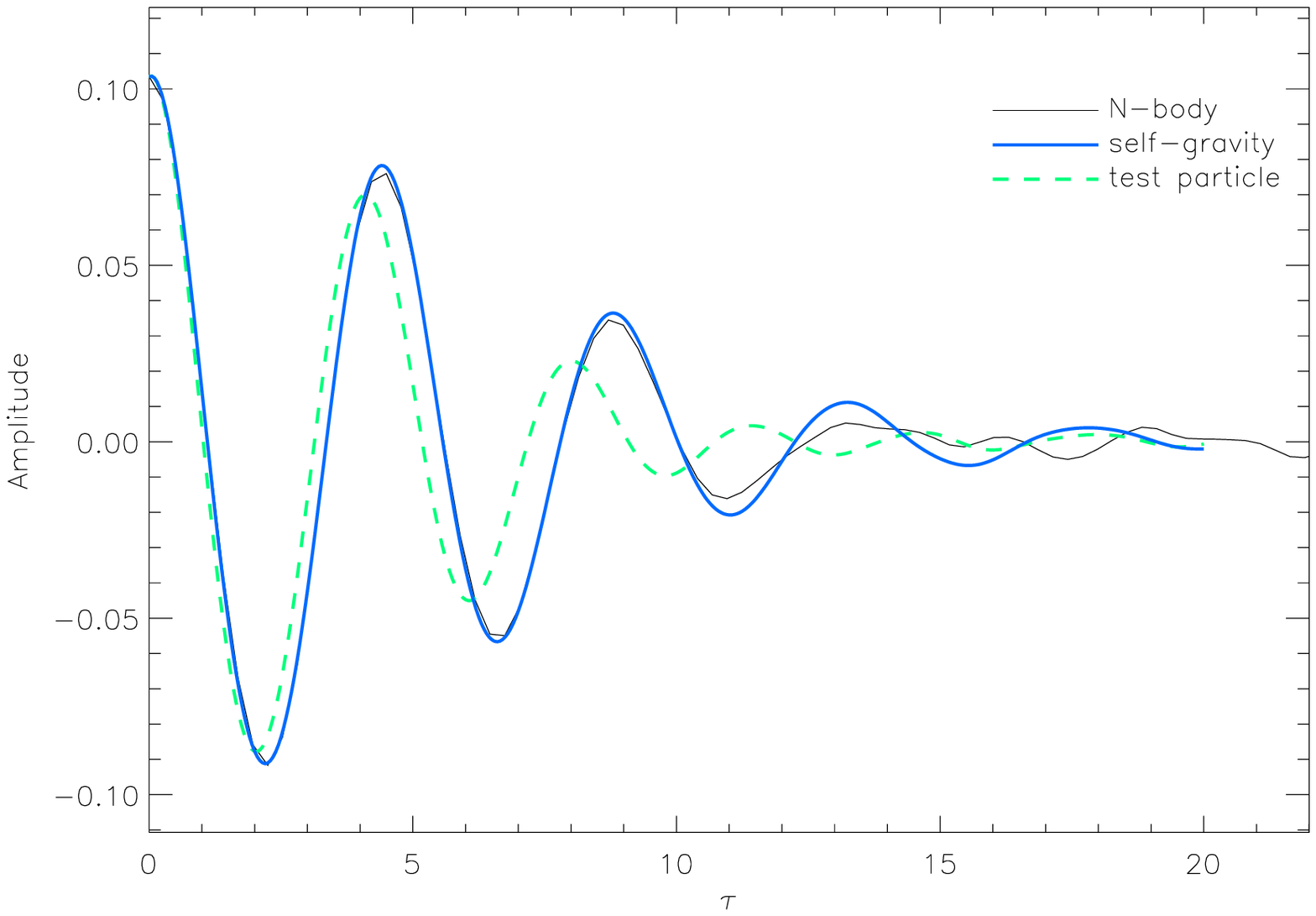}}
\caption{Comparison of test-particle (thick dashed line), linear
self-gravitating (thick solid line), and $N$-body simulation (thin
solid line) coefficient evolutions for a model with $c_{1,1}(0)=0.1$.
The frequency of the linear analysis agrees with the $N$-body
simulation.  As predicted by the simple system analyzed in
Section~\ref{sols}, this frequency is less than the test-particle
frequency.} 
\label{comptestlinearN}
\end{figure}

Figure~\ref{compevlp} shows the early evolution of a few low-order
coefficients for a self-gravitating system.  As in the test-particle
case, the (1,1) term is initially perturbed and the line styles are
the same as in Figure~\ref{compevtp}.  The overall behavior is
strikingly similar to that of the test-particle case. The difference
between the coefficient evolution equation and the $N$-body
simulations is shown in Figure~\ref{diffevlp} for $c_{1,1}(0)=0.1$.
Figure~\ref{comptestlinearN} compares the time behavior of $c_{1,1}$
calculated from Equation~\ref{cbelp} for both the test-particle and
linear self-gravitating cases. Also shown is the corresponding
self-gravitating $N$-body simulation for $c_{1,1}(0)=0.1$.  As
predicted by the analysis of the $2\times 2$ system, the oscillation
period of the linear self-gravitating curve is longer than that of the
test-particles curve, and agrees with the $N$-body simulation.

\begin{figure}
\scalebox{0.5}{
\includegraphics{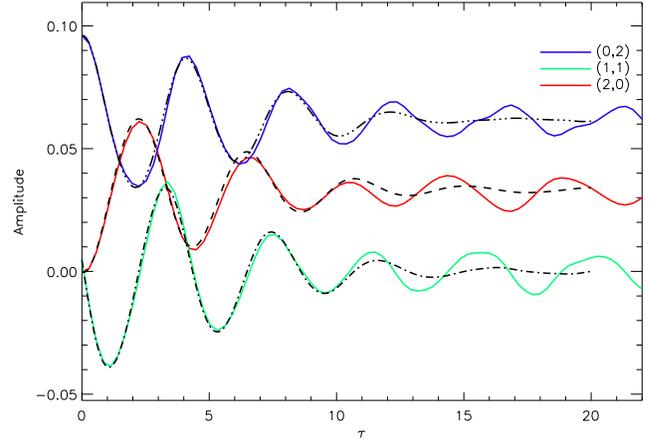}}
\caption{Linear perturbation coefficient evolutions for an initial
(0,2) perturbation with strength $b=0.1$.  The line styles are
the same as in Figure~\ref{compevlp}.  The significant difference
between these two figures is that the presence of a time-independent
solution is evident as the (0,2) and (2,0) coefficients approach
constant, non-zero values with increasing time.}
\label{tind}
\end{figure}

We have also constructed systems with initial (0,2) and (2,0)
perturbations for comparison.  Figure~\ref{tind} highlights the fact
that (0,2) linear perturbations lead to time-independent solutions, as
the initial amplitude in the (0,2) term decreases to a non-zero value
as the (2,0) amplitude grows to a non-zero value [similar results
develop given an initial (2,0) perturbation].  This behavior is
reminiscent of the results of the simple coefficient system developed
in Section~\ref{sols}.

As is evident in Figures~\ref{diffevtp} and~\ref{diffevlp},
differences between $\left< c_{m,n} \right>$ and $c_{m,n}$  grow with
time in both test-particle and self-gravitating cases.  To test
whether this difference is due to the truncation of the polynomial
expansion, we have also solved the equations on a grid where $m_{\rm
max}$ and $n_{\rm max}$ are twice the values used to calculate the
curves in the figures above.  The rms differences between the two
$c_{m,n}$ are an order of magnitude smaller than the differences
between $\left< c_{m,n} \right>$ and $c_{m,n}$ , so the root cause of
the $N$-body/coefficient dynamics discrepancy lies in $N$.  As the
number of particles in a simulation is increased from $N=1024$ to
$N=2048$, the differences between $\left< c_{m,n} \right>$ and
$c_{m,n}$ diminish.  We have created two self-gravitating ensembles
with initial (1,1) perturbations and $b=0.1$.  One ensemble is formed
from realizations using $N=1024$ particles, the other uses $N=2048$.
We have calculated the ratios between the rms $N$-body/coefficient
dynamics differences and the error in the mean values for the
ensemble.  For the $N=1024$ ensemble, these ratios are 1.08, 0.70, and
0.89 for the (1,1), (0,2), and (2,0) terms, respectively.  For the
$N=2048$ ensemble, the error in the mean values decrease by
$\sqrt{2}$, but the rms $N$-body/coefficient dynamics differences
decrease by a larger factor, producing ratios of 0.97, 0.66, and 0.83
for the (1,1), (0,2), and (2,0) terms, respectively.  We interpret
these results to mean that the coefficient dynamics solutions
represent the truly collisionless behavior of the system  for small
enough perturbations, which the $N$-body simulations can only produce
in the limit $N \rightarrow \infty$.

In addition to the finite $N$ effects described above, $N$-body
simulations always contain some degree of non-linearity that is not
included in first-order perturbation theory.  We have investigated the
effects of non-linearity by tracking ensemble coefficient evolutions
as perturbation strengths is increased.  Specifically, $N=1024$
particle evolutions with initial (0,2), (1,1), and (2,0) perturbations
have been performed, each with strengths $b=0.1$, $b=0.2$, and
$b=0.3$.  Table~\ref{btable} quantifies the impact of perturbation
strength.  As one would expect, the growth of the difference ratios
indicates the increased presence of non-linear effects in the $N$-body
simulations.  Depending on a particular tolerance for discrepancies,
one could set a limit on the maximum perturbation strength allowable
for an evolution to remain in the linear regime.

\begin{table*}
\begin{minipage}{12cm}
\caption{Ratio of the average rms difference of coefficients,
$c_{m,n}$, calculated from Equation~\ref{cbelp} and the ensemble
values $\left< c_{m,n} \right>$ of the nonlinear $N$-body simulations
to the error in the mean of $\left< c_{m,n} \right>$.  Each horizontal
row has a different initially perturbed polynomial, and each column
lists this ratio for a different $c_{m,n}$ for different perturbation
strength $b$.\label{btable}}
\begin{tabular}{|c||c|c|c||c|c|c||c|c|c|}
\hline
 & \multicolumn{3}{c||}{$c_{0,2}$ ratio} &
\multicolumn{3}{c||}{$c_{1,1}$ ratio} & \multicolumn{3}{c|}{$c_{2,0}$
ratio} \\ \cline{2-10}
\raisebox{2.3ex}[0pt]{initial term} & $b=0.1$ & $b=0.2$ & 
$b=0.3$ & $b=0.1$ & $b=0.2$ & $b=0.3$ & 
$b=0.1$ & $b=0.2$ & $b=0.3$\\ \hline
\hline
$c_{0,2}$ & 0.75 & 1.63 & 3.55 & 1.14 & 1.69 & 3.86 & 1.02 & 
1.70 & 3.89 \\ \hline
$c_{1,1}$ & 0.70 & 1.02 & 2.04 & 1.08 & 1.30 & 1.90 & 0.89 & 
1.18 & 1.73 \\ \hline
$c_{2,0}$ & 1.10 & 2.21 & 4.54 & 1.89 & 3.75 & 7.38 & 1.59 & 
3.13 & 6.28 \\ \hline
\end{tabular}
\end{minipage}
\end{table*}

\section{Discussion}\label{discuss}

We have demonstrated that  a set of orthonormal polynomial terms based
on the equilibrium distribution function is useful for investigating
the evolution of one-dimensional, self-gravitating, collisionless
systems,  at least for small linear perturbations from equilibrium.
The polynomial coefficients interact via diagonal-neighbor couplings,
producing an alternate view of the evolution of these systems in terms
of coefficients $c_{m,n}$ on the $(m,n)$ grid. 

From a simple ``wall clock'' point-of-view, a coefficient evolution
takes less time to perform than an ensemble calculation.  For example,
our midpoint method solution (with the timestep discussed in
Section~\ref{sims}) on an $m_{\rm max}=60$, $n_{\rm max}=60$ grid
takes approximately twenty minutes to evolve to $\tau=30$ on a 3GHz
microprocessor using a translator language like MatLab or IDL.   On
the same machine, each equivalent self-gravitating $N=1024$
realization of an ensemble takes approximately three minutes to evolve
with efficient compiled Fortran code.  The 100 realization ensembles
discussed in this work would have taken five hours to complete, if run
serially.   The computation time dramatically increases for larger
particle numbers, mainly because of the shorter crossing times in the
heap-sort algorithm.  Each $N=2048$ realization takes approximately
eight times as long as the $N=1024$ version.   Beyond the time issue,
the coefficient evolution approach allows one to follow the behavior
of a perfectly smooth distribution function, instead of using density
histograms as one must do in the $N$-body approach. 

Having said that, coefficient evolutions exhibit their own
shortcomings.  The size of the $c_{m,n}$ grid that is used determines
the resolution of the features of $(x,v)$ phase that can be
represented.  While we have not developed an analytical relationship
between resolution scale and boundary size, there is a conceptual
connection that is useful.  If one thinks of the Hermite polynomials
used to describe velocity behavior, there are $m_{\rm max}$ roots for
the $H_{m_{max}}(\varpi)$ term.  Phase-space features that have velocity
scales smaller than the spacing between the roots will not be properly
captured by terms with $m < m_{\rm max}$ and go unresolved by the
coefficient dynamics approach.  A similar argument may be made for the
position aspect of any phase-space feature.  For any given situation,
one must make a decision regarding an acceptable level of phase-space
resolution and then choose $m_{\rm max},n_{\rm max}$ appropriately, a
familiar situation when dealing with classical phase-space
\citep{ll51}.

For a constant, uniform time-step, the stability of the midpoint
method places a rather severe requirement on the size of $d\tau$.
Instabilities occur in the coefficients at large $m$ and $n$ where the
derivatives are largest.  We have found that $d\tau \lesssim
(10\sqrt{m_{max}}n_{max})^{-1}$ for the stability of the midpoint
method. As a result, the number of calculations required for an
$n_{max}\times n_{max}$ grid scales like $n_{max}^{7/2}$. In
principle, one could use different time steps for different regions of
phase space, but we did not explore this strategy, since the savings
would be modest.

\begin{figure}
\scalebox{0.65}{
\includegraphics{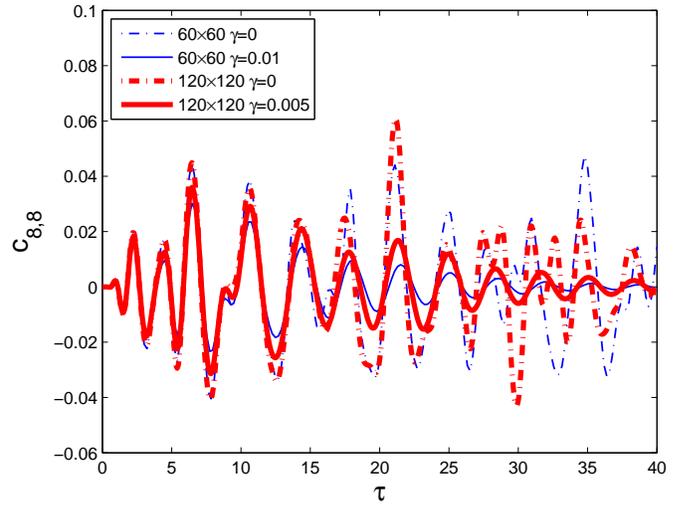}}
\caption{Evolution of $c_{8,8}$ given an initial $c_{1,1}$
perturbation, highlighting boundary effects and damping. The
dot-dashed curves represent non-damped solutions defined on different
fixed-boundary grids; thin -- $60\times 60$ grid and thick -- $120
\times 120$.  The two agree up to $\tau\approx 20$ where reflection
from the grid boundary starts to affect the $60\times 60$ curve. Also
shown are a damped ($\gamma=1/100$) solution on a $60\times 60$ grid
(thin solid curve), and a damped ($\gamma=1/200$) solution on a $120
\times 120$ grid (thick solid curve).  This form of damping causes the
distribution function to decay before it reaches the boundaries, but
affects low-order coefficients minimally.}
\label{c22damping}
\end{figure}

The biggest shortcoming of our approach, however, is the fact that
boundary conditions of the grid limit the time that
Equation~\ref{cbelp} can be used to accurately calculate low-order
coefficients so that they can be considered effectively those of an
unbounded system.  The simple ``fixed'' boundary conditions we have
used, where $c_{m_{max},n}= c_{m, n_{max}}=0$, reflect coefficient
amplitude ``waves'' incident on the boundary back into the interior,
where they propagate back toward the low-order coefficients near
$(0,0)$.  The effect of this reflection at long times on the evolution
$c_{2,2}$ is clearly evident in Figure~\ref{c22damping}.  

To test the impact of the reflection on the coefficient evolution we
have compared two grids, one with twice the dimensions of the other.
While a strict quantitative measure is difficult, our investigations
suggest that terms with $m,n \lesssim 10$ calculated on an $m_{\rm
max}=n_{\rm max}=60$ grid are in good agreement with the same terms
calculated on an $m_{\rm max}=n_{\rm max}=120$ grid for $\tau \lesssim
20$.  Terms with smaller $m,n$ values show the best agreement between
the two grid evolutions.  It must also be remembered that only a
finite range of $m,n$ terms can be accurately tracked during an
$N$-body evolution.  For the $N$-values adopted in this work, we have
found that, even with ensemble averaging, the particle noise level is
comparable to the amplitude of $m,n \gtrsim 10$ terms that arise.
While it may be possible to design non-reflecting boundaries (as one
would do for the ordinary wave equation), we have not been able to
find a scheme that is stable and provides significant improvement over
fixed boundary conditions. 

\begin{figure}
\scalebox{0.65}{
\includegraphics{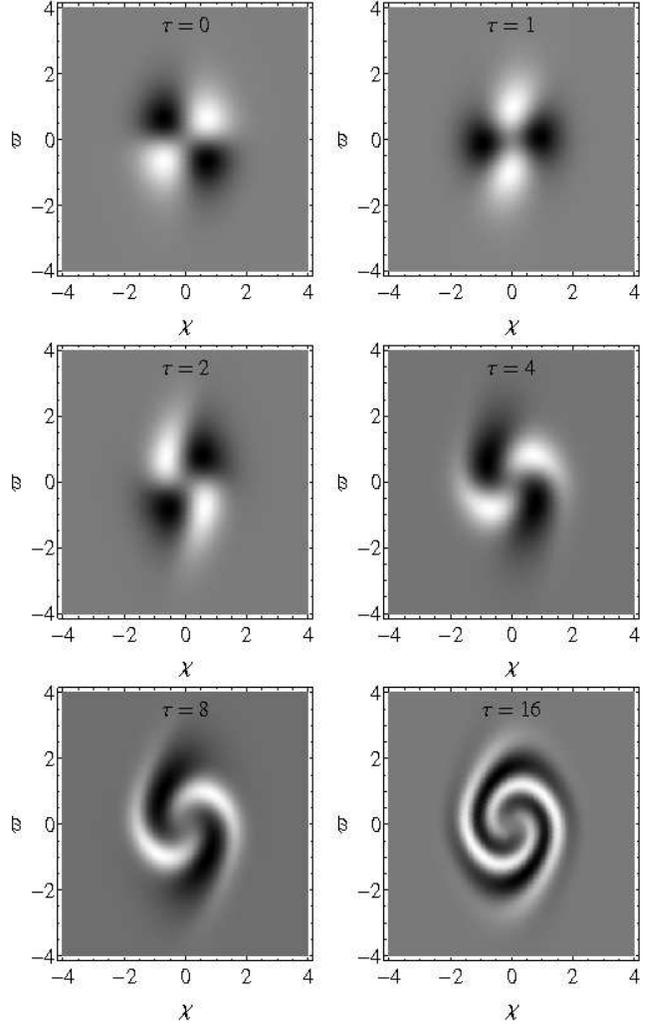}}
\caption{Phase-space evolution of an initial (1,1) perturbation in a
self-gravitating system, created from evolving coefficients according
to Equation~\ref{cbelp} and reconstructing the phase-space distribution
function from Equation~\ref{fsum}.  The coefficient evolution uses a
$120\times120$ grid, with a damping factor $1/\gamma=200$ to eliminate
distortions of higher modes.}
\label{xvt}
\end{figure}

Another strategy that shows promise is to include a damping term in
the evolution equations.  Instead of simply propagating toward large
$(m,n)$, disturbances also decay so that the boundaries of the grid
are never reached, avoiding distortions due to reflection. Several
varieties of damping term have been tested, and the most appealing has
the form,
\begin{eqnarray}
\dot{c}_{m,n}^{damping}
& = & -\gamma\left(|L_{m,n}^{m-1,n-1}|+|L_{m,n}^{m+1,n-1}|+ \right.
\nonumber \\
& & \left. |L_{m,n}^{m-1,n+1}|+|L_{m,n}^{m+1,n+1}|\right)c_{m,n},
\end{eqnarray}
where $1/\gamma$ can be interpreted as a quality factor for the
coefficient dynamics.  This form has the advantage that the damping
rate is proportional to the derivatives of the coefficients, and,
other than a gradual decay of the disturbance, its impact on the
dynamics of low-order coefficients seems to be minimal. In particular,
the oscillation frequencies are hardly affected, for small damping.
Figure~\ref{c22damping} shows the time-behavior of the $c_{8,8}$
amplitude calculated with two different damping factors.  In contrast
to the effects of boundary reflections, which lead to complicated
distortions throughout the ($m,n$) grid, the effect of damping seems
to be a simpler filtering of the amplitude.  However, even with this
simple form of damping it is not easy to obtain a formula for the
region of validity of a coefficient evolution for a given $\tau$.  The
simplest approach is to pick a grid size and then find the minimum
$\gamma$ that keeps the perturbation from reaching the boundaries.
Figure~\ref{xvt} shows a reconstructed phase space plot for
$m_{max}=n_{max}=120$ with a quality factor $1/\gamma=200$.

Regardless of the numerical technique adopted, this polynomial
expansion analysis of the Vlasov equation provides a novel, and
useful, view of the behavior of one-dimensional self-gravitating
systems.  While not in the scope of this introductory work, one can
imagine several directions any future investigations using this
analysis might take. For example one might study the aftermath of the
interactions of multiple isolated systems, investigating collisionless
processes in the non-equilibrium remnant.  Additionally, determining
the stationary states of one-dimensional systems or the frequency
spectrum of the $L$ matrix could also be approached.  One could extend
the analysis to second-order to investigate the onset of nonlinear
effects, like stability or chaotic behavior.  The nearest-neighbor
coupling of the coefficients leads to ``local'' continuity-type
dynamics of conserved quantities like energy and fine-grained entropy
on the $(m,n)$ grid that should give further insight into the
non-equilibrium thermodynamics of these systems. 

\section*{Acknowledgements}

E.\ I.\ Barnes acknowledges support from a Research Infrastructure
Program grant from the Wisconsin Space Grant Consortium.  The authors
thank the anonymous referee for several suggestions that have led to a
more robust publication.

\end{document}